\documentstyle[amssymb,aps,12pt]{revtex}

\begin{document}
\draft
\author{J. Matyjasek$^{1}$ and O. B. Zaslavskii$^{2}$}
\address{$^{1}$Institute of Physics, Maria Curie-Sklodowska University, \\
pl. Marii Curie-Sklodowskiej 1, 20-031 Lublin,\\
Poland\\
E-mails: matyjase@tytan.umcs.lublin.pl,\\
jurek@iris.umcs.lublin.pl\\
$^{2}$Department of Physics, Kharkov V.N. Karazin's National University,\\
Svoboda Sq.4,\\
Kharkov 61077, Ukraine\\
E-mail: aptm@kharkov.ua}
\title{Quantum backreaction of massive fields and self-consistent semiclassical
extreme black holes and acceleration horizons }
\maketitle

\begin{abstract}
We consider the effect of backreaction of quantized massive fields on the
metric of extreme black holes (EBH). We find the analytical approximate
expression for the stress-energy tensor for a scalar (with an arbitrary
coupling), spinor and vector fields near an event horizon. We show that,
independent of a concrete type of EBH, the energy measured by a freely
falling observer is finite on the horizon, so that quantum backreaction is
consistent with the existence of EBH. For the Reissner-Nordstr\"{o}m EBH
with a total mass $M_{tot}$ and charge $Q$ we show that for all cases of
physical interest $M_{tot}<Q$ . We also discuss different types of
quantum-corrected Bertotti-Robinson spacetimes, find for them exact
self-consistent solutions and consider situations in which tiny quantum
corrections lead to the qualitative change of the classical geometry and
topology. In all cases one should start not from a classical background with
further adding quantum corrections but from the quantum-corrected
self-consistent geometries from the very beginning.
\end{abstract}



\section{introduction}

Nowadays, the physical relevance and importance of the issue of extreme
black holes (EBH) does not need detailed clarification. Let us only mention
briefly such issues as the end-point of black hole evaporation, information
loss, the black hole entropy, etc. In fact, the background of EBH can serve
as a promising testing area of potential predictions of (yet not
constructed) quantum gravity in the semiclassical domain. Meanwhile,
recently, the very fact of existence of semiclassical black holes became the
subject of discussion \cite{lowe}. In this paper Lowe presented strong
arguments confirming the existence of semiclassical EBH. These arguments,
however, are of phenomenological nature in that they tacitly assume that the
components of the stress-energy tensor (SET) of quantized field and their
relevant combinations with the metric functions remain finite on the
horizon. Meanwhile, this is not obvious in advance. For example, in the
background of the Reissner-Nordstr\"{o}m (RN) EBH this fact for massless
radiation was established only by virtue of thorough numerical calculations 
\cite{anders}.

In such a situation it looks reasonable to elaborate the general
backreaction approach to EBH similar to that \cite{york} applied to
Schwarzschild black holes. However, the attempt of moving in this direction
immediately encounters the following difficulty which reveals the crucial
difference between the nonextreme and extreme black holes in the given
context. In the first case, it was sufficient to choose the fixed background
and carry out calculations perturbatively, whereas in the second one the
very nature of the background becomes not trivial. Say, for the classical
Reissner-Nordstr\"{o}m (RN) with charge $Q$ and mass $M$ minor changes
around the extreme relationship $M=Q$ can convert EBH to the nonextreme hole
or naked singularity. Correspondingly, one should be very careful in
examining changes caused by quantum effects at the border of so different
kinds of spacetimes. Therefore, the accent in the backreaction program for
EBH (at least, on the first step) is to be shifted as compared to the
Schwarzschild case: first of all, it is necessary to elucidate, whether or
not EBH are compatible with backreaction. It looks natural to take a generic
EBH\ metric, ``dressed'' by surrounding quantum fields, and elucidate,
whether or not backreaction is compatible with the property that the Hawking
temperature $T_{H}=0$. In turn, this invokes information about the SET of
quantized field spacetime of a generic spherically-symmetrical EBH. For
massless fields, this task is extremely difficult. Meanwhile, for massive
fields recent progress in deriving general expressions for SET \cite{matrn}, 
\cite{infeld} makes the task tractable. As the calculation of SET is the key
to the problem of existence of quantum-corrected EBH, let us dwell upon this
issue in a more detail.

\section{General features of SET\ of massive fields in curved manifolds}

According to a standard viewpoint, the renormalized stress--energy tensor
(SET) of quantized fields evaluated in appropriate state encodes all
available information of quantum field theory in curved background, and
(beside the classical part) it serves as a source term of the semiclassical
Einstein field equations. Unfortunately, mathematical complexities prevent
exact analytical treatment and in most of physically interesting situations
it cannot be expressed in terms of known special functions. Moreover, what
is of principal interest in further applications is not the SET itself
evaluated in the particular geometry, but rather its functional dependence
on a wide class of metrics. Therefore, we are confronted with two serious
problems: construction of the SET on the one hand, and studying the effect
of the quantized field upon the spacetime geometry on the other. It is
natural therefore, that to address these problems, at least partially, one
should employ approximate methods.

It seems that for the massive fields in the large mass limit considered in
this paper, an approximation based on the DeWitt-Schwinger expansion is of
required generality, allowing, in principle, to attack the problem of
backreaction prerturbatively. Moreover, in some situations (also considered
here), it is even possible to construct {\it exact} solutions of the
semiclassical field equations, or, what is more common, guided by physical
considerations, guess the appropriate form of the line element. Although
such a procedure is limited to especially simple geometries with a high
degree of symmetries, obtained results are of particular interest and
importance.

For the massive fields in a curved spacetime, the renormalized effective
action, $W_{R}$, constructed by means of the DeWitt--Schwinger method is
given by 
\begin{equation}
W_{R}=\frac{1}{32\pi ^{2}m^{2}}\int d^{4}xg^{1/2}\sum_{n=3}^{\infty }\frac{%
(n-3)!}{\left( m^{2}\right) ^{n-2}}\left[ a_{n}\right] ,  \label{act}
\end{equation}
where $\left[ a_{n}\right] $ is the coincidence limit of the n-th
Hadamard-DeWitt coefficient and $m$ is the mass of the field, and the first
three terms of the DeWitt-Schwinger expansion have been absorbed by
quadratic terms of the generalized classical gravitational action in the
process of the renormalization of the bare constants. As the complexity of
the Hadamard-DeWitt coefficients rapidly grows with $n$, the practical use
of (\ref{act}) is confined to the first order of $W_{R}$, which involves the
integrated coincidence limit of fourth Hadamard--DeWitt coefficient $a_{3}$
computed by Gilkey \cite{gilkey}. Being constructed from local, geometrical
quantities, the first order effective action does not describe the process
of particle creation which is a nonlocal phenomenon, however, for
sufficiently massive fields, the contribution of real particles may be
neglected and the DeWitt-Schwinger $W_{R}$ satisfactorily approximates the
total effective action. It could be shown that for massive scalar, spinor,
and vector fields the first-order effective action could be compactly
written in a form \cite{avra} 
\begin{eqnarray}
W_{ren}^{(1)}\, &=&\,{\frac{1}{192\pi ^{2}m^{2}}}\int d^{4}xg^{1/2}\left(
c_{1}^{(s)}R\Box R\,+\,c_{2}^{(s)}R_{\mu \nu }\Box R^{\mu \nu
}\,+\,c_{3}^{(s)}R^{3}\,+\,c_{4}^{(s)}RR_{\mu \nu }R^{\mu \nu }\right.
\label{w1} \\
&+&\,c_{5}^{(s)}RR_{\mu \nu \rho \sigma }R^{\mu \nu \rho \sigma
}\,+\,c_{6}^{(s)}R_{\nu }^{\mu }R_{\rho }^{\nu }R_{\mu }^{\rho
}\,+\,c_{7}^{(s)}R^{\mu \nu }R_{\rho \sigma }R_{~\mu ~\nu }^{\rho ~\sigma
}+c_{8}^{(s)}R_{\mu \nu }R_{\lambda \rho \sigma }^{\mu }R^{\nu \lambda \rho
\sigma }\,  \nonumber \\
&+&\left. \,\,c_{9}^{(s)}{R_{\rho \sigma }}^{\mu \nu }{R_{\mu \nu }}%
^{\lambda \gamma }{R_{\lambda \gamma }}^{\rho \sigma
}\,+\,c_{10}^{(s)}R_{~\mu ~\nu }^{\rho ~\sigma }R_{~\lambda ~\gamma }^{\mu
~\nu }R_{~\rho ~\sigma }^{\lambda ~\gamma },\right) =\,{\frac{1}{192\pi
^{2}m^{2}}}\sum_{i=1}^{10}c_{i}^{(s)}W_{i},  \nonumber
\end{eqnarray}
where the numerical coefficients depending on the spin of the field are
listed in Table I. As the approximate stress--energy tensor is obtained by
functional differentiation of (\ref{w1}) with respect to the metric tensor, 
\begin{equation}
T^{\mu \nu (q)}=\frac{1}{96\pi ^{2}m^{2}g^{1/2}}\sum_{i=1}^{10}c_{i}^{(s)}%
\frac{\delta }{\delta g_{\mu \nu }}W_{i}\text{,}  \label{qtens}
\end{equation}
one concludes that within the framework of the adopted approximation, it is
expressed as a linear combination of purely geometrical terms with the
numerical coefficients depending on the spin of the field, and consequently
is independent of boundary conditions. Since the calculations are carried
out for the Euclideanized geometry, the resulting Green functions bear close
relations with the temperature Green functions, and in the black hole
spacetimes in the absence of the superradiant modes, the thus obtained SET
may be interpreted in terms of the thermal state. An alternative approach,
consisting in the construction of appropriate Green functions by summing
(integrating) WKB approximants of the mode functions of the scalar field
equation with arbitrary coupling to a curvature, has been proposed in Ref. 
\cite{ahs}. It has been shown that, to obtain the lowest order terms in the
DeWitt--Schwinger expansion, one has to employ the results of 6-th order
WKB. Moreover, detailed analyses, both analytical and numerical, of the
stress--energy tensor of the quantized massive scalar field carried out in
the Reissner--Nordstr\"{o}m spacetime confirmed that the DeWitt--Schwinger
approximation yields reasonable results as long as the mass of the field is
sufficiently large\cite{ahs}. Specifically, it was shown that for quantized
scalars in the vicinity of the event horizon of RN black hole, the
approximation remains within a few percents of the exact (numerical) value
if the condition $mM\geq 2$ holds.

General expressions for the first nonvanishing order of SET of the massive
scalar, spinor, and vector fields, which generalize earlier results of
Frolov and Zel'nikov for vacuum type-D geometries \cite{FZ} , are
constructed in \cite{matrn} and \cite{infeld}. They may be, in principle,
used in any spacetime provided the temporal changes of the background are
slow and the ratio of the Compton length to the characteristic radii of
curvature are small. However, because of computational complexity, their
practical use is limited to simple spacetimes. Happily, there are
considerable simplifications for the class of metrics considered in this
paper: spherically-symmetric geometries with vanishing curvature scalar and
the spacetimes with maximally symmetric subspaces. On the other hand,
however, in some physically important and computationally tractable cases,
as for example Kerr or Kerr--Newman spacetimes, there are superradiant
modes, and the SET constructed along the lines of the DeWitt--Schwinger
approximation must be interpreted with care. However, in spite of its
inherent limitations, the DeWitt-Schwinger method is still the most general
one not restricted to any particular type of symmetry.

In this paper we shall use the general results of \cite{matrn} and \cite
{infeld} to evaluate the renormalized SETs of the massive scalar, spinor,
and vector field in the spacetime of extremal black holes (EBH). The
calculations for a general metric turn out to be extremely complicated and
are of little practical use. Fortunately, for the issue of the existence of
quantum-corrected EBH and the properties of corresponding self-consistent
solutions of the Einstein equations, it is sufficient to expand the metric
potentials in the vicinity of the event horizon into the Taylor series and
examine SET constructed for this simplified line element. Additionally, we
shall construct and examine the SET in the Bertotti - Robinson-like
spacetimes obtained by expanding the near--horizon geometry into a whole
manifold.

\section{semiclassical extreme black holes}

\subsection{quantum backreaction and degenerate horizon}

The metric under consideration reads 
\begin{equation}
ds^{2}=-Udt^{2}+V^{-1}dr^{2}+r^{2}d\Omega ^{2}\text{,}  \label{m}
\end{equation}
where the form of $V(r)$ can be found from $00$ component of the Einstein
equations. It is equal to 
\begin{eqnarray}
V &=&1-\frac{r_{+}}{r}-\frac{2\tilde{m}(r)}{r}\text{,}  \label{v} \\
\tilde{m}(r) &=&4\pi \int_{r_{+}}^{r}dr^{\prime }r^{^{\prime }2}\rho
(r^{\prime })\text{;}  \nonumber
\end{eqnarray}
where $\rho =-T_{0}^{0}$, the SET $T_{\mu }^{\nu }=T_{\mu }^{\nu
(cl)}+T_{\mu }^{\nu (q)}$. Here the first term comes from a classical
source, the second one is due to the contribution of quantum fields and is
to be understood as a quantum average with respect to the the Hartle-Hawking
state, renormalized in a proper way. Let us assume that the role of a
classical source is played by an electromagnetic field ($T_{\mu }^{\nu
(cl)}\equiv T_{\mu }^{\nu (em)}$), so we deal with the quantum-corrected RN
black hole. Correspondingly, $\tilde{m}(r)=m_{em}+m_{q}$, where $m_{em}=%
\frac{Q^{2}}{2}(\frac{1}{r_{+}}-\frac{1}{r})$, $m^{q}=4\pi
\int_{r_{+}}^{r}dr^{\prime }r^{^{\prime }2}\rho ^{q}(r^{\prime })$. Here it
is implied that the event horizon is located at $r=r_{+}$. In this sense $%
r_{+}$ is ``exact'' value of the horizon radius (to some extent the word
``exact'' is conditional since $T_{\mu }^{\nu (q)}$ is known in the one-loop
approximation only). The function $U=Ve^{2\psi }$, where the concrete form
of the function $\psi (r)$ can be found from the $(rr)$ Einstein equation.
Then the Hawking temperature 
\begin{equation}
T_{H}=\frac{V^{^{\prime }}(r_{+})}{4\pi }e^{\psi (r_{+})}\text{.}  \label{h}
\end{equation}
The explicit form of $V$ is

\begin{eqnarray}
V &=&1-\frac{2m(r)}{r}+\frac{Q^{2}}{r^{2}}\text{,}  \label{v2} \\
m(r) &=&\tilde{m}(r)+\frac{Q^{2}}{2r}+\frac{r_{+}}{2}=M+m^{q}(r)\text{.} 
\nonumber
\end{eqnarray}

It follows from the definitions that $m^{q}(r_{+})=0$ (no room for
radiation) and $m(r_{+})=M$. The condition that $r_{+}$ is the root of $%
V(r)=0$ means that 
\begin{equation}
g(r)\equiv r^{2}V=r^{2}-2m(r)r+Q^{2}=0\text{.}  \label{r}
\end{equation}

If $\psi (r_{+})$ is bounded on a horizon, the answer to the question
whether or not \ a black hole can reach the extreme state is determined by
whether or not $V^{\prime }(r_{+})$ can turn into zero. In fact, the
finiteness of $\psi (r_{+})$ on the horizon, typical of a nonextreme black
hole, becomes a nontrivial issue in the extreme case. It is equivalent to
the problem whether or not the energy measured by a freely falling observer
remains finite on the horizon (see, cf. \cite{anders}). Thorough numerical
calculations showed that $\psi (r_{+})$ is indeed finite for the RN EBH in
the case when quantized fields are massless \cite{andrn}. The behavior of $%
\psi $ near the horizon is one of key issues examined below in the present
paper for the case of massive fields. Let us, however, put for a moment this
matter aside and assume that $\psi (r_{+})$ is indeed finite.

The equation (\ref{r}) should be satisfied at $r=r_{+}$ independently of
whether the horizon is extremal. Here $m(r)$ is an unknown function but near 
$r_{+}$ we can expand it like $m(r)=M+A(r-r_{+})+...$, where $A=4\pi
r_{+}^{2}\rho ^{q}(r_{+})$ has the order $\varepsilon\, =\,\hbar/M^{2}$. It
is more convenient to write: 
\begin{equation}
m(r)=M_{o}+Ar+...\text{,}  \label{a}
\end{equation}
where $M_{0}=M-Ar_{+}$. Now we get the equation 
\begin{equation}
g(r)=r^{2}(1-2A)-2rM_{0}+Q^{2}=0\text{,}  \label{m0}
\end{equation}
whence the roots are 
\begin{equation}
r_{\pm }=\frac{M_{0}}{1-2A}\pm \sqrt{(\frac{M_{0}}{1-2A})^{2}-\frac{Q^{2}}{%
1-2A}}\text{.}  \label{rr}
\end{equation}
If one adjust parameters in such a way that $M_{0}^{2}=Q^{2}(1-2A)$, then 
\begin{equation}
r_{+}=r_{-}=Q/\sqrt{1-2A}=\frac{M}{1-A}  \label{dr}
\end{equation}
is the the double root of the function $g(r)$ and $M=\frac{M_{0}(1-A)}{1-2A}=%
\frac{Q(1-A)}{\sqrt{1-2A}}$.

From physical grounds, it is essential for the existence of EBH that the 
{\it total energy }density on the horizon (including, in the RN case,
electromagnetic contribution) be positive (and large enough), so the sign of
quantum contribution itself (if it is not too large) is not so crucial.

Let us also describe another, seemingly ``obvious'' approach that, however,
contains a hidden trap. In other (less successfull) notations one can write
the identity that follows from substitution $r=r_{+}$ into $g(r_{+})=0$: 
\begin{equation}
r_{+}^{2}-2Mr_{+}+Q^{2}=0\text{,}  \label{un}
\end{equation}
whence 
\begin{equation}
r_{+}=M+\alpha \sqrt{M^{2}-Q^{2}\text{,}}  \label{al}
\end{equation}
$\alpha =1$ $\ $or $\alpha =-1$. At the first glance, the choice $\alpha =1$
should correspond to the event horizon as it is the case for classical RN
black holes. However, when one insert $r=r_{+}$ into the identity $%
g(r_{+})=0 $, it turns out that, as a matter of fact, one inserts into an
equation its own root as a parameter of this very equation (by contrast, $%
M_{0}$, $Q$ and $A$ are independent parameters and the roots $r_{\pm }$ are
expressed in their terms directly according to (\ref{rr})). This procedure
is not quite safe and can lead to the appearance of ``spurious'' solutions 
\cite{lowe}. Therefore, one should verify its self-consistency to make sure
that the root under consideration is a ``true'' one.

It is instructive to demonstrate this explicitly. Let us compare two
different presentation of the same quantity - (\ref{rr}) and (\ref{al}).
Then after some manipulations we have 
\begin{equation}
\alpha \sqrt{M^{2}-Q^{2}}=\sqrt{Z}+\frac{M_{0}A}{1-2A}\text{,}  \label{z}
\end{equation}
where $Z=(\frac{M_{0}}{1-2A})^{2}-\frac{Q^{2}}{1-2A}$. If $A>0$, one should
take $\alpha =1$ (it is supposed that $A$ is not too large; in fact, $A\ll 1$%
). However, near the extreme state $Z\rightarrow 0$ and $A<0$, one should
choose $\alpha =-1$.

Thus, quantum backreaction shifts \ the double root to a new position but
does not change its character qualitatively \cite{lowe}. In so doing,
however, in a sharp contrast with the classical case, the horizon for $A<0$
lies at $r_{+}=M-\sqrt{M^{2}-Q^{2}}$ .

\subsection{General approach to extreme black holes dressed by quantized
massive fields}

Meanwhile, as is mentioned above, the fact that backreaction leaves the
possibility for the existence of double root of eq. (\ref{m0}) is
insufficient in itself for making conclusions about the existence of extreme
quantum-corrected black holes. According to $rr$ and $tt$ components of the
Einstein equations, 
\begin{equation}
\psi =4\pi \int_{\infty }^{r}drF(r)\text{, }F(r)=r\frac{T_{1}^{1}-T_{0}^{0}}{%
V}\text{.}  \label{11}
\end{equation}
Here the lower limit of integration is set to infinity since it is supposed
that spacetime infinity is flat and $\psi =0$. Below we will consider the
case of massive fields only for which $T_{\mu }^{\nu }\rightarrow 0$ as $%
r\rightarrow \infty $ (for massless fields it is assumed usually that a
system is enclosed into a finite cavity, otherwise $T_{\mu }^{\nu
}\rightarrow const\neq 0$).

The key question is whether or not the quantity $F$ is finite on the
horizon. Further details depend on the possibility of power expansion of the
metric near the horizon. As far as massless fields are concerned, the
counterpart from two-dimensional black hole physics shows \cite{triv} that
(i) for a generic fixed metric the function $F(r)$ diverges on the horizon
that indicates the qualitative change of the metric of extreme black hole
under influence of quantum field, (ii) if, instead of fixing a metric in
advance, one chooses it as a {\it self-consistent }solution of field
equations with backreaction taken into account, the existence of an extreme
black hole is compatible with quantum backreaction but (iii) the power
expansion of the metric near the horizon fails to be analytic. On the other
hand, numerical calculations for the four-dimensional RN background \cite
{andrn} showed that $F$ remains finite on the horizon.

There are the following subtleties in our problem. As, by assumption,
quantum field is neutral, it does not screen a charge, so its value $Q$ is
the same for an original classical and quantum-corrected backgrounds. Then
it follows from (\ref{rr}) that $r_{\pm }<Q$, if $\rho ^{q}(r_{+})<0$. Had
we chosen the classical background with such a relationship between
parameters and tried to take into account quantum backreaction by building
up the perturbation series, we would have obtained physically meaningless
result. Indeed, if the root of the equation $g(r)=0$ is less than $Q$, it
corresponds in a classical language to the Cauchy (not the event) horizon,
where the stress-energy tensor of quantum fields is known to blow up. This
obstacle testifies clearly that, instead of using an standard scheme (pure
classical background plus perturbative quantum corrections) we should start
from the {\it self-consistent quantum-corrected background from the very
beginning}.

Our strategy consists in the following. As the issue of the existence of
extreme black holes (EBH) demands knowledge of behavior of the metric near
the horizon only, let us consider the vicinity of the horizon of a generic
EBH, expand the metric near the horizon into the power series and examine,
whether or not the quantity $F$ remains finite on the horizon. If $F$ is
finite (that means the finiteness of SET in the orthonormal reference frame
of a free falling observer \cite{anders}), one obtains the power expansion
for the SET too, so the full self-consistent solution can be obtained by a
direct expansion into the Taylor series with respect to $r-r_{+}$.

For EBH the conjectured power expansion in terms of $r-r_{+}$ looks like $%
V=a(r-r_{+})^{2}+b(r-r_{+})^{3}+...$ For concrete calculations it is more
convenient, however, to use, instead of $r$, the proper distance $l$ from
some fixed point. Then we have $\frac{dl}{dr}=-\frac{1}{\sqrt{V}}$.
Substituting into this equation the power expansion for $V$ near the
horizon, we find 
\begin{equation}
r-r_{+}=A_{1}e^{-l/\rho }+A_{2}e^{-2l/\rho }+...\text{.,}  \label{rcoord}
\end{equation}
where $\rho =a^{-1/2}$ and the integration constant $l_{0}$ is absorbed by
coefficients according to $A_{1}=r_{+}\exp (l_{0}/\rho )$, $%
A_{2}=r_{+}^{2}b/a\exp (2l_{0}/\rho )$.

We can write down 
\begin{equation}
ds^{2}=-dt^{2}U(l)+dl^{2}+r^{2}(l)d\Omega ^{2}\text{.}  \label{l}
\end{equation}
In what follows we assume that power expansion of $U$ in terms of $r-r_{+}$
starts from the terms of $(r-r_{+})^{2}$, as it is typical for EBH. In term
of $l$ this function reads 
\begin{equation}
U=e^{-2l/\rho }f(l)\text{, \quad }f=f_{0}+f_{1}e^{-l/\rho }+f_{2}e^{-2l/\rho
}\text{.}  \label{near}
\end{equation}
The expressions for SET of massive fields in the metric (\ref{near}) are
very cumbersome. However, what is the most important for us, it is their
general structure. It turns out that near the horizon 
\begin{equation}
F=F_{0}+F_{1}e^{-l/\rho }+F_{2}e^{-2l/\rho }+...  \label{F}
\end{equation}
with {\it finite} coefficients $F_{i}$ ($i=0,1,2...$)$.$ The expressions for
the components of SET read 
\begin{equation}
T_{\mu }^{\nu }=t_{\mu }^{\nu (0)}+t_{\mu }^{\nu (1)}e^{-l/\rho }+t_{\mu
}^{\nu (2)}e^{-2l/\rho }+...\text{,}  \label{set}
\end{equation}
where explicit expression for the coefficients $t_{\mu }^{\nu }$ are listed
in Appendix for different kinds of field. It is essential that in all cases
it turns out that $t_{0}^{0(0)}=t_{1}^{1(0)}$ and $t_{0}^{0(1)}=t_{1}^{1(1)}$
that just leads to the finiteness of $F$ on the horizon.

It is worth stressing that the finiteness of $F$ for the case of massive
fields is shown for any EBH irrespective of whether or not its metric obeys
the system of field equations and the type of the theory to which these
field equation correspond. Now this general result is applied for the most
physically interesting case of RN EBH, dressed by its quantum radiation.

\subsection{quantum-corrected RN extreme black hole}

From physical viewpoint, it is natural to fix the total mass measured by a
distant observer at infinity (the microcanonical boundary condition). Then
we have from (\ref{v2}) 
\begin{equation}
M_{tot}=M+m^{q}\text{, }m^{q}=-4\pi \int_{r_{+}}^{\infty }drr^{2}T_{0}^{0(q)}%
\text{, }M=\frac{r_{+}^{2}+Q^{2}}{2r_{+}}\text{.}  \label{mass}
\end{equation}

The condition of extremality $V^{\prime }(r_{+})=0$ entails 
\begin{equation}
r_{+}^{2}(1-2A)=Q^{2}\text{.}  \label{ext}
\end{equation}
In all cases $m^{q}=\alpha _{s}m^{-2}r_{+}^{-3}$, where $\alpha _{0}=-\tilde{%
\alpha}\frac{17}{441}$, $\alpha _{1/2}=-\tilde{\alpha}\frac{19}{147}$, $%
\alpha _{1}=-\tilde{\alpha}\frac{107}{441},$ and $\tilde{\alpha}=\frac{1}{%
720\pi }$. From (\ref{mass}), (\ref{ext}) it is seen that the correction of
the first order in $M$ cancel and we obtain 
\begin{equation}
M_{tot}=Q+\alpha _{s}m^{-2}r_{+}^{-3}\text{.}  \label{tot}
\end{equation}
In the main approximation $\rho =r_{+}=M=Q$. We see that for all physically
relevant cases $\alpha _{s}<0$ and $M_{tot}<Q$. Thus, a distant observer
measuring by precise devices the total mass and charge of an extreme RN
black hole, had he relied on classical notions only and neglect quantum
backreaction completely, would have been led to the wrong conclusion that in
fact the object under investigation is rather naked singularity than a black
hole. In other words, the quantum-corrected solution of Einstein-Maxwell
equations under discussion not only acquires some small corrections from
backreaction of quantum fields but resides in a pure quantum domain where
the existence of classical black holes (both extremal or non-extremal) is
strictly forbidden.

One can also find the quantum-corrected position of the horizon in terms of
physical parameters. Taking into account (\ref{dr}) and noticing that $%
T_{0}^{0(q)}(r_{+})=\eta _{s}m^{-2}r_{+}^{-6}$, $\eta _{s}=\frac{\mu _{s}}{%
2880\pi ^{2}}$, $\mu _{0}=\frac{16}{21}-4(\xi -1/6)$, $\mu _{1/2}=\frac{37}{%
14}$, $\mu _{1}=\frac{114}{7}$ \cite{matrn}, we obtain $A=4\pi \eta
_{s}m^{-2}r_{+}^{-4}$, $r_{+}=Q(1-2A)^{-1/2}\approx Q(1+A)$. Equivalently,
in terms of a total mass, $r_{+}=M_{tot}(1+\beta _{s}m^{-2}M_{tot}^{-4})$, $%
\beta _{s}=4\pi \eta _{s}-\alpha _{s}$. Here $\beta _{0}=\tilde{\alpha}\left[
\frac{353}{441}-4(\xi -\frac{1}{6})\right] $, $\beta _{1/2}=\tilde{\alpha}%
\frac{815}{294}$, $\beta _{1}=\tilde{\alpha}\frac{7289}{441}$.

\section{quantum-corrected BR-like spacetimes}

\subsection{Self-consistent solutions without cosmological term}

It is obvious that it is impossible to find exact solution of backreaction
equation for realistic four-dimensional EBH in all space and this is the
reason why we were forced to restrict ourselves by the treatment of the
vicinity of the horizon only. Meanwhile, there exists another class of
objects for which exact solutions (in the one-loop approximation) can indeed
be found - metrics with acceleration horizons (Bertotti-Robinson (BR)
spacetime and its modifications). Such spacetimes have topology $(r,t)\times
S_{2}$, where $S_{2}$ is two-dimensional sphere, so that the coefficient
standing at the angular part of a line element is constant. The physical
relevance of such spacetimes stems, in particular, from the fact that it can
serve as approximation to the true metric of EBH in the vicinity of the
horizon. Apart from this, such a kind of a metric appears in the limiting
transition from nonextreme black holes to extreme ones \cite{prl}, \cite{prd}%
, \cite{accel}. SET for the BR spacetime was studied in \cite{kof1}, \cite
{matrn}. Now, however, we start not from the BR itself, but from its
quantum-corrected version.

The general form of metrics under consideration is 
\begin{equation}
ds^{2}=-U(l)dt^{2}+dl^{2}+r_{0}^{2}d\Omega ^{2}\text{,}  \label{gbr}
\end{equation}
where it is assumed that there exists a horizon on which $U\rightarrow 0$.
In the coordinates ($x^{1}$, $\theta $, $\phi $, $t$) the SET\ of
electromagnetic field is 
\begin{equation}
8\pi T_{\mu }^{\nu (em)}=\frac{Q^{2}}{r_{0}^{4}}(-1\text{, }1\text{, }1\text{%
, }-1)\text{.}  \label{em}
\end{equation}

However, the expression for SET of quantized fields in such a background is
rather complicated and will not be written here. Fortunately, if we restrict
ourselves to the BR-like spacetime and guess its quantum-corrected version,
we obtain a very simple answer in a compact form. We found that this
procedure is tractable for the following cases:

Metric $BR1$: 
\begin{equation}
ds^{2}=-dt^{2}\rho ^{2}sh^{2}\frac{l}{\rho }+dl^{2}+r_{0}^{2}d\Omega ^{2}%
\text{,}  \label{ro}
\end{equation}

\begin{equation}
G_{\mu }^{\nu }=(-\frac{1}{r_{0}^{2}}\text{, }\frac{1}{\rho ^{2}}\text{, }%
\frac{1}{\rho ^{2}}\text{, }-\frac{1}{r_{0}^{2}})\text{,}  \label{ei}
\end{equation}

Metric $BR2$:

\begin{equation}
ds^{2}=-dt^{2}\exp (-2l/\rho )+dl^{2}+r_{0}^{2}d\Omega ^{2}  \label{br2}
\end{equation}
The Einstein tensor has the same form (\ref{ei}).

Metric $dS_{2}\times S_{2}$:

\begin{equation}
ds^{2}=-dt^{2}\sigma ^{2}\sin ^{2}\frac{l}{\sigma }+dl^{2}+r_{0}^{2}d\Omega
^{2}\text{,}  \label{br3}
\end{equation}
\begin{equation}
G_{\mu }^{\nu }=(-\frac{1}{r_{0}^{2}}\text{, }-\frac{1}{\sigma ^{2}}\text{,}-%
\text{ }\frac{1}{\sigma ^{2}}\text{, }-\frac{1}{r_{0}^{2}})  \label{eisin}
\end{equation}

Metric $Rindler_{2}\times S_{2}$: 
\begin{equation}
ds^{2}=-dt^{2}l^{2}+dl^{2}+r_{0}^{2}d\Omega ^{2}\text{.}  \label{ri}
\end{equation}
\begin{equation}
G_{\mu }^{\nu }=(-\frac{1}{r_{0}^{2}}\text{, }0\text{, }0\text{ }-\frac{1}{%
r_{0}^{2}})  \label{eir}
\end{equation}

In all cases indicated above SET of quantum fields can be written as 
\begin{equation}
8\pi T_{\mu }^{\nu (q)}=C(f_{1}\text{, }f_{2}\text{, }f_{2}\text{, }f_{1})%
\text{,}  \label{q}
\end{equation}
where $f_{1}$ , $f_{2}$ are simple constants depending on curvatures of the
two dimensional maximally-symmetric subspaces, $C=\frac{1}{12\pi ^{2}m^{2}}$%
, and $m$ is a mass of a field. The form of (\ref{q}) follows from the fact
that for (\ref{ro}), (\ref{br2}), (\ref{br3}), and (\ref{ri}) the covariant
derivatives of the Riemann tensor and its contractions vanish that
considerable simplifies the SET given by (\ref{qtens}).

The fact that $T_{2}^{2(q)}=T_{3}^{3(q)}$ is a simple consequence of
symmetry of the metric with respect to rotations. The equality $%
T_{0}^{0(q)}=T_{1}^{1(q)}$ can be understood as follows: the metrics of the
type (\ref{gbr}) can be obtained as a result of certain limiting transition
from black holes ones, in the process of which the near-horizon geometry
expands into a whole manifold \cite{accel}; then SET pick up their values
from the horizon where the regularity condition demands just the validity of
this equality.

It turns out that for all cases the function $f_{1}$ and $f_{2}$ share the
common general structure. In $BR1$ and $BR2$ $f_{1}=\rho
^{-6}r_{0}^{-6}(a_{1}r_{0}^{6}+b_{1}r_{0}^{4}\rho ^{2}+c_{1}\rho ^{6})$, $%
f_{2}=\rho ^{-6}r_{0}^{-6}(a_{2}\rho ^{6}+b_{2}\rho
^{4}r_{0}^{2}+c_{2}r_{0}^{6})$, where the coefficients $a_{i}$, $b_{i}$ and $%
c_{i}$ ($i=1$, $2$) are the same for both metrics.

For all values of the spin $a_{2}=-a_{1}$, $b_{2}=-b_{1}$, $c_{2}=-c_{1}$.

In the scalar case ($s=0$):

$a_{1}=\frac{1}{105}(8-84\xi +420\xi ^{2}-840\xi ^{3})$, $b_{1}=-\frac{1}{105%
}(7-112\xi +630\xi ^{2}-1260\xi ^{3})$,

$c_{1}=\frac{1}{105}(4-42\xi +210\xi ^{2}-420\xi ^{3})$;

$s=1/2$: $a_{1}=\frac{20}{420}$, $b_{1}=\frac{7}{420}$, $c_{1}=\frac{10}{420}
$;

$s=1$: $a_{1}=\frac{8}{35}$, $b_{1}=\frac{7}{35}$, $c_{1}=\frac{4}{35}$.

Metric $dS_{2}\times S_{2}$ (now $f_{1}=\sigma
^{-6}r_{0}^{-6}(a_{1}r_{0}^{6}+b_{1}r_{0}^{4}\sigma ^{2}+c_{1}\sigma ^{6})$
and $f_{2}=\sigma ^{-6}r_{0}^{-6}(a_{2}\sigma ^{6}+b_{2}\sigma
^{4}r_{0}^{2}+c_{2}r_{0}^{6})$):

$a_{2}=a_{1}$, $b_{2}=b_{1}$, $c_{2}=c_{1}$.

$s=0$: $a_{1}=\frac{1}{105}(-8+84\xi -420\xi ^{2}+840\xi ^{3})$, $b_{1}=%
\frac{1}{105}(-7+112\xi -630\xi ^{2}+1260\xi ^{3})$, $c_{1}=\frac{1}{105}%
(4-42\xi +210\xi ^{2}-420\xi ^{3})$;

$s=1/2$: $a_{1}=-\frac{20}{420}$, $b_{1}=\frac{7}{420}$, $c_{1}=\frac{10}{420%
}$;

$s=1$: $a_{1}=-\frac{8}{35}$, $b_{1}=\frac{7}{35}$, $c_{1}=\frac{4}{35}$.

Metric $Rindler_{2}\times S_{2}$:

$f_{1}=ar_{0}^{-6}$, $f_{2}=-2ar_{0}^{-6}$, $a_{0}=\frac{8-63\eta -3780\eta
^{3}}{945}$ ($\eta =\xi -1/6$), $a_{1/2}=\frac{1}{42}$, $a_{1}=\frac{4}{35}$.

In the limits $\rho \rightarrow \infty $ and $\sigma \rightarrow \infty $
the metrics $BR1$and $dS_{2}\times S_{2}$ turn into $Rindler_{2}\times S_{2}$%
. One can check that in these limits the function $f_{1}$ and $f_{2}$ go
smoothly to their values for the metric $Rindler_{2}\times S_{2}$.

Let the classical background be of the $BR1$ or $BR2$ type metrics (The
metric (\ref{br3}) cannot appear on the pure classical level without a
cosmological constant). Then we obtain two independent equations from the
Einstein ones: 
\begin{equation}
-\frac{1}{r_{0}^{2}}=-\frac{Q^{2}}{r_{0}^{4}}+Cf_{1}  \label{1}
\end{equation}
\begin{equation}
\frac{1}{\rho ^{2}}=\frac{Q^{2}}{r_{0}^{4}}+Cf_{2}  \label{2}
\end{equation}

Taking the sum of (\ref{1}) and (\ref{2}) and noticing that $f_{1}+f_{2}=%
\frac{(r_{0}^{2}-\rho ^{2})}{r_{0}^{6}\rho ^{6}}\chi $, where $\chi $ has
the structure $\chi =\alpha _{1}r_{0}^{4}+\alpha _{2}\rho ^{4}+\alpha
_{3}\rho ^{2}r_{0}^{2}$ ($\alpha _{i}$ are pure numbers), we obtain 
\begin{equation}
(r_{0}^{2}-\rho ^{2})(1-\frac{C\chi }{r_{0}^{4}\rho ^{4}})=0\text{.}
\label{dif}
\end{equation}
Now take into account that $C\sim \lambda _{PL}^{2}\lambda ^{2}$, where $%
\lambda =m^{-1}$ is the Compton length and $\lambda _{PL}$ is the Planckian
length. Then a simple estimate shows that the second factor in (\ref{dif})
can turn into zero (provided the proper signs appear in it) for $r_{0}\ll
\lambda $ only, so far beyond the region of validity of WKB approximation.
Therefore, we will not discuss such a possibility further and assume that
there is only one root of eq. (\ref{dif}): $r_{0}=\rho $. This means that BR
spacetime remains exact solution of semiclassical equations (cf. \cite{kof1}%
, \cite{solod}, \cite{accel}). Making use of eq. (\ref{1}) and writing $%
f_{1}(r_{0}=\rho )=\gamma r_{0}^{-6}$, where $\gamma =a_{1}+b_{1}+c_{1}$, we
find that $r_{0}^{2}=Q^{2}-C\gamma r_{0}^{-2}$, whence, in the same
approximation, 
\begin{equation}
r_{0}^{2}=Q^{2}(1-C\gamma Q^{-4})\text{,}  \label{re}
\end{equation}

In (\ref{re}) the second term in parenthesis represents only a small
correction but account for this correction can be crucial in the following
sense. Let $\gamma >0$. Then we have $Q>r_{0}$. Let us proceed, for
definiteness, in the canonical ensemble approach in which a charge (rather,
than a potential on the boundary) should be fixed. First, if $Q\neq r_{0}$,
the classical metric with an acceleration horizon of the type (\ref{1}) or (%
\ref{br2}) is impossible at all. Instead of it, we would have a geometry of
a RN black hole. Second, for $Q>r_{0}$ with $r_{0}$ being the horizon
radius, we would have, moreover, a naked singularity. It is clear that the
procedure in which the ground state is chosen as a classical geometry with a
naked singularity with quantum corrections, calculated on such a background
perturbatively, is physically unacceptable. Instead of it, we should from
the very beginning use the quantum-corrected geometry and check the
condition of self-consistency for the corresponding parameters. In our case
it is possible due to a relative simplicity of the BR geometry that enables
us to find SET just not only for a classical BR itself but also for
quantum-corrected version of it (below we will see that it is also the case
even if the quantum-corrected metrics changes its form - for example, due to
the cosmological term).

The value of $\gamma $ for different values of field spin (subscript
indicates the value of a spin $s$): $\gamma _{0}=\frac{5-14\xi }{105}$, $%
\gamma _{1}=\frac{19}{35}$, $\gamma _{1/2}=\frac{37}{420}$. Thus, for a
spinor and vector field $\gamma >0$ as well as for the scalar case with the
minimal and conformal coupling.

\subsection{Nonzero cosmological constant}

Now for the $BR1$ and $BR2$ metrics field equations read 
\begin{equation}
\frac{1}{r_{0}^{2}}=\frac{Q^{2}}{r_{0}^{4}}-\Lambda -Cf_{1}  \label{l1}
\end{equation}
and 
\begin{equation}
\frac{1}{\rho ^{2}}=\frac{Q^{2}}{r_{0}^{4}}+\Lambda +Cf_{2}  \label{l2}
\end{equation}
For the $BR3$ case we have, instead of (\ref{l2}), 
\begin{equation}
\frac{1}{\sigma ^{2}}=-\frac{Q^{2}}{r_{0}^{4}}-\Lambda -Cf_{2}\text{.}
\label{13}
\end{equation}

Then it follows from eq. (\ref{l1}) that 
\begin{equation}
\frac{1}{r_{0}^{2}}=\frac{1}{2Q^{2}}\pm \frac{1}{Q}\sqrt{\frac{1}{4Q^{2}}%
+\Lambda +Cf_{1}}\text{.}  \label{root}
\end{equation}
The term with $Cf_{1}$ represents a small correction to classical
quantities, so that in the expression for $f_{1}$ one can replace $r_{0}$
and $\rho $ by the classical values obtained for $C=0$ . If $\Lambda >0$,
only the solution with the $+$ sign should be taken. We will discuss the
more interesting for us case $\Lambda =-\left| \Lambda \right| <0$. Let $%
\Lambda $ be very close to the value $\Lambda _{0}=-\frac{1}{4}Q^{-2}$ for
which the radical in (\ref{root}) turns into zero. In $f_{1}$we can put in
the main approximation $r_{0}=2^{1/2}Q$, $\rho =\infty $, neglecting
corrections of the order $C.$ Classically, we would have the product of
two-dimensional Rindler and sphere (\ref{ri}), for which, according to the
above results, $f_{1}$ $=\bar{a}Q^{-6}$, $f_{2}=-2f_{1}$ with $\bar{a}=\frac{%
1}{8}a>0$ for spinor and vector fields as well as for the scalar case both
for the conformal and minimal coupling. If $\Lambda -\Lambda _{0}<0$, the 
{\it classical} constant curvature solutions of the type (\ref{1}) do not
exist at all. However, if the difference $\Lambda -\Lambda _{0}$ is very
small and such that $\Lambda -\Lambda _{0}+C\bar{a}Q^{-6}>0$, the solutions (%
\ref{root}) do exist. Let us substitute the expression for $r_{0}^{2}$ into
eq. (\ref{l2}) for $\rho $. Then 
\begin{equation}
\frac{1}{\rho ^{2}}=\frac{1}{r_{0}^{2}}+2\Lambda +C(f_{1}+f_{2})=2(\Lambda
-\Lambda _{0})+C(f_{1}+f_{2})\pm \frac{1}{Q}\sqrt{\Lambda -\Lambda
_{0}+Cf_{1}}\text{.}  \label{rol}
\end{equation}
Consider two cases.

a) $\Lambda =\Lambda _{0}$. Then we obtain two solutions. The first one is 
\begin{equation}
\rho ^{2}=\frac{Q^{4}}{\sqrt{C\bar{a}}}\text{,}  \label{r0e}
\end{equation}
where the term $-\bar{a}Q^{-6}$ has been dropped. Thus, quantum corrections
force the geometry to switch from (\ref{ri}) to (\ref{ro}) or (\ref{br2}).
The second solution is formally complex. In fact, this means that instead of
(\ref{ro}) we have the geometry (\ref{br3}). Now $\sigma ^{2}=\frac{Q^{4}}{%
\sqrt{C\bar{a}}}$.

2) $\Lambda =\Lambda _{0}-Cf_{1}=\Lambda _{0}-C\bar{a}Q^{-6}$. Then we have
the metric (\ref{br3}) with parameters 
\begin{equation}
r_{0}^{2}=2Q^{2}\text{, }\sigma ^{2}=\frac{1}{C(f_{1}-f_{2})}=\frac{Q^{6}}{3C%
\bar{a}}\text{,}  \label{1sol}
\end{equation}
where we took into account the expressions for the metric $Rindler_{2}\times
S_{2}.$

\section{concluding remarks}

We have considered EBH in equilibrium with quantized massive fields and
demonstrated that for {\it any} EBH the components of SET, measured by a
free-falling observer, remain finite. If a metric obeys the Eisntein
equations, this entails that semiclassical EBH do exist as their
self-consistent solutions. The key point of our treatment consisted in
restricting to the analysis of the near-horizon geometry that enabled us to
avoid the complexity connected with obvious impossibility to find explicit
self-consistent solutions in the whole domain.

We considered also BR-like spacetime, closely connected to the issue of EBH,
and have shown that quantum-corrected BR remain to be the exact solutions of
one-loop field equations. In so doing, the relationship between parameters
of solutions can be such that classically they are absent at all and only
quantum effects make their existence (for fixed values of these parameters)
possible. Apart from this, near some critical points in the space of
solutions tiny quantum corrections can lead to the change of the type of the
BR spacetime, the scale of curvature remaining pure classical. Thus, quantum
corrections not only shift slightly the values of relevant physical
quantities but lead to qualitative changes in the geometry and topology.

The questions about near-horizon behavior of SET and self-consistent EBH for
massless fields as well as properties of self-consistent BR spacetimes
deserve separate treatment.

\section{Acknowledgment}

O. Z. thanks Claus Kiefer, Paul Anderson and David Lowe for stimulating
discussions and correspondence on quantum-corrected extreme black holes. O.
Z. also thanks for hospitality Department of Physics of Freiburg University
(where the preliminary stage of this work was started due to kind support of
DAAD) and Institute of Physics of Maria Curie-Sklodowska University in
Lublin, where the most part of this work has been done.

The authors are grateful to Sergey Sushkov for useful correspondence on BR
spacetimes.

\appendix

\section{Power expansion for T$_{\protect\mu }^{\protect\nu (q)}$ of massive
fields near the horizon of a generic ebh}

In this Appendix we collect a number of formulas for the components of SET
of the massive scalar, spinor, and vector fields in the vicinity of the
event horizon of a generic eternal black hole, which are used in this paper.
Functionally differentiating $W_{i}$ with respect to the metric tensor,
performing the necessary symmetrizations and simplifications, and inserting
thus obtained results into (\ref{qtens}) one obtains the general form of the
renormalized SET of the quantized massive fields. As the resulting formulas
are rather complicated, we shall not display them here, and a reader is
referred to \cite{matrn} and \cite{infeld} for further details.
Subsequently, constructing the components of Riemann tensor, its
contractions and necessary covariant derivatives for the line element (\ref
{l}), inserting thus obtained results into the general expressions of SET 
\cite{infeld}, combining them with the appropriate spin-dependent numerical
coefficients $c_{i}^{(s)}$, and finally making use of the explicit form of $%
U(l)$ and $r(l)$ as given by (\ref{near}) and (\ref{rcoord}), and collecting
the terms with the like powers of $z=$ $e^{-l/\rho }$, one has 
\begin{equation}
T_{\mu }^{\nu (q)}=\sum_{i=0}t_{\nu }^{\nu (i)}z^{i}=\frac{1}{96\pi ^{2}m^{2}%
}\sum_{i=0}\widetilde{t}_{\nu }^{\nu (i)}z^{i}\text{.}  \label{a2}
\end{equation}
Although the near-horizon power expansions of the line element (\ref{l})
look rather simple, the complexity of the SET rapidly increases with the
order of expansion, practically invalidating calculations of $t_{\nu }^{\nu
(i)}$ for $i\geq 3$. Below the results for $i=0,1$ are listed.

Closer analysis of the coefficients $c_{i}^{(0)}$ given in Table I indicates
that the general SET is a third-order polynomial in $\eta =\xi -1/6$, with
coefficients given by a purely local, geometrical terms.

\begin{equation}
\widetilde{t}_{0}^{0(0)}=\widetilde{t}_{1}^{1(0)}={\frac{8}{945}}\,{\frac{%
2\,r_{+}^{6}+\rho ^{6}}{\rho ^{6}\,r_{+}^{6}}}-{\frac{\eta \left( \rho
^{6}-\rho r_{+}^{4}+2r_{+}^{6}\right) }{15\rho ^{6}r_{+}^{6}}}\,-4\,{\frac{%
\eta ^{3}\,(2\,r_{+}^{6}-3\,\rho ^{2}\,r_{+}^{4}+\rho ^{6})}{\rho
^{6}\,r_{+}^{6}}}\text{,}  \label{t00}
\end{equation}
\begin{eqnarray}
\widetilde{t}_{0}^{0(1)} &=&\widetilde{t}_{1}^{1(1)}=-{\frac{16}{315}}\,{%
\frac{A_{1}\,(r_{+}^{6}+\rho ^{6})}{r_{+}^{7}\,\rho ^{6}}}+{\frac{2}{15}}\,{%
\frac{\eta \,{A_{1}}\left( 3\rho ^{6}-\rho ^{4}r_{+}^{2}-\rho
^{2}r_{+}^{4}+3r_{+}^{6}\right) }{\rho ^{6}r_{+}^{7}}}  \nonumber \\
&&+24\,{\frac{\eta ^{3}\,A_{1}\,(\rho ^{6}+r_{+}^{6}-\rho
^{2}\,r_{+}^{4}-r_{+}^{2}\,\rho ^{4})}{r_{+}^{7}\,\rho ^{6}}}\text{,}
\end{eqnarray}
\begin{equation}
\widetilde{t}_{2}^{2(0)}=\widetilde{t}_{3}^{3(0)}=\,{\frac{4\eta
^{3}(\,r_{+}^{6}+2\,\rho ^{6}-3\,r_{+}^{2}\,\rho ^{4})\,}{r_{+}^{6}\,\rho
^{6}}}+\,{\frac{\eta \left( \rho ^{6}-\rho ^{4}r_{+}^{2}+r_{+}^{6}\right) }{%
15\rho ^{6}r_{+}^{6}}}-{\frac{2}{945}}\,{\frac{(8\,\rho ^{6}+4\,r_{+}^{6})}{%
r_{+}^{6}\,\rho ^{6}}}\text{,}
\end{equation}
\begin{eqnarray}
\lefteqn{\widetilde{t}_{2}^{2(1)}=\widetilde{t}_{3}^{3(1)}=\frac{2A_{1}}{315}%
\,\frac{16\,\rho ^{6}\,\,+17\,\,r_{+}^{6}\,+7\,\,r_{+}^{2}\,\rho ^{4}}{%
\,r_{+}^{7}\,\rho ^{6}}}  \nonumber \\
&&-\frac{\eta \,}{30f_{0}\rho ^{6}r_{+}^{7}}(24A_{1}f_{0}\rho
^{6}+20A_{1}f_{0}\rho ^{4}r_{+}^{2}+3f_{1}\rho
^{4}r_{+}^{3}+40A_{1}f_{0}\rho ^{2}r_{+}^{4}  \nonumber \\
&&+100A_{1}f_{0}r_{+}^{6}+15f_{1}r_{+}^{7})+{\frac{2\eta ^{2}}{%
f_{0}\,r_{+}^{5}\,\rho ^{6}}}(4\,A_{1}\,f_{0}\,\rho
^{4}+3\,f_{1}\,r_{+}^{3}\,\rho ^{2}  \nonumber \\
&&+16\,A_{1}\,r_{+}^{4}\,f_{0}+16A_{1}\,r_{+}^{2}\,f_{0}\,\rho
^{2}+6\,f_{1}\,r_{+}^{5})-\frac{6\eta ^{3}}{{f}_{0}\,r_{+}^{7}\,\rho ^{6}}%
(20\,A_{1}\,r_{+}^{2}\,f_{0}\,\rho ^{4}  \nonumber \\
&&+8\,\rho ^{6}\,A_{1}\,f_{0}+8\,A_{1}\,r_{+}^{4}\,f_{0}\,\rho
^{2}-36A_{1}\,r_{+}^{6}\,f_{0}+3\,f_{1}\,r_{+}^{3}\,\rho ^{4}  \nonumber \\
&&+12f_{1}\,r_{+}^{5}\,\rho ^{2}-15\,f_{1}\,r_{+}^{7})\text{,}
\end{eqnarray}

Repeating the calculations with the coefficients $c_{i}^{(1/2)}$ one obtains

\begin{equation}
\widetilde{t}_{0}^{0(0)}=\widetilde{t}_{1}^{1(0)}=\frac{1}{420}\,\frac{%
7\,r_{+}^{4}\,\rho ^{2}+10\,\rho ^{6}+20\,r_{+}^{6}}{r_{+}^{6}\,\rho ^{6}}%
\text{,}
\end{equation}
\begin{equation}
\widetilde{t}_{0}^{0(1)}=\widetilde{t}_{1}^{1(1)}=-\frac{1}{210}\,\frac{%
A_{1}\,(30\,\rho ^{6}+7\,r_{+}^{2}\,\rho ^{4}+7\,r_{+}^{4}\,\rho
^{2}+30\,r_{+}^{6})}{r_{+}^{7}\,\rho ^{6}}\text{,}
\end{equation}
\begin{equation}
\widetilde{t}_{2}^{2(0)}=\widetilde{t}_{3}^{3(0)}=-\frac{1}{420}\,\frac{%
20\,\rho ^{6}+7\,r_{+}^{2}\,\rho ^{4}+10\,r_{+}^{6}}{r_{+}^{6}\,\rho ^{6}}
\end{equation}
\begin{equation}
\widetilde{t}_{2}^{2(1)}=\widetilde{t}_{3}^{3(1)}=-{\frac{1}{840}}\,{\frac{%
21\,f{_{1}}\,r_{+}^{3}\,\rho ^{4}-136\,A_{1}\,r_{+}^{6}\,f_{0}-240\,\rho
^{6}\,A_{1}\,f_{0}-140\,A_{1}\,r_{+}^{2}\,f_{0}\,\rho ^{4}}{%
f_{0}\,r_{+}^{7}\,\rho ^{6}}}\text{,}
\end{equation}

Finally, for the massive vector fields

\begin{equation}
\widetilde{t}_{0}^{0(0)}=\widetilde{t}_{1}^{1(0)}=\frac{1}{35}\,\frac{%
8\,r_{+}^{6}+7\,\rho ^{2}\,r_{+}^{4}+4\,\rho ^{6}}{\rho ^{6}\,r_{+}^{6}}%
\text{,}
\end{equation}
\begin{equation}
\widetilde{t}_{0}^{0(1)}=\widetilde{t}_{1}^{1(1)}=-\frac{2}{35}\,\frac{%
A_{1}\,(12\,r_{+}^{6}+7\,\rho ^{2}\,r_{+}^{4}+12\,\rho
^{6}+7\,r_{+}^{2}\,\rho ^{4})}{r_{+}^{7}\,\rho ^{6}}\text{,}
\end{equation}
\begin{equation}
\widetilde{t}_{2}^{2(0)}=\widetilde{t}_{3}^{3(0)}=-\frac{1}{35}\,\frac{%
7\,r_{+}^{2}\,\rho ^{4}+8\,\rho ^{6}+4\,r_{+}^{6}}{\rho ^{6}\,r_{+}^{6}}%
\text{,}
\end{equation}
\begin{eqnarray}
\widetilde{t}_{2}^{2(1)} &=&\widetilde{t}_{3}^{3(1)}=-\frac{1}{%
210f_{0}r_{+}^{7}\rho ^{6}}(-280\,A_{1}\,r_{+}^{4}\,f_{0}\,\rho
^{2}+63\,f_{1}\,r_{+}^{3}\,\rho ^{4}-420\,A_{1}\,r_{+}^{2}\,f_{0}\,\rho ^{4}
\nonumber \\
&&-124\,A_{1}\,r_{+}^{6}\,f_{0}-288\,\rho ^{6}\,A_{1}\,f_{0}+105\,{f_{1}}%
\,r_{+}^{5}\,\rho ^{2})\text{.}
\end{eqnarray}
The third-order coefficients of the expansion (\ref{a2}), $\widetilde{t}%
_{\mu }^{\nu (2)}$, are too lengthty to be presented here. On the other
hand, however, of principal importance in the analyses of the regularity of $%
F$ is the difference between $(tt)$ and $(rr)$ components of the
stress-energy tensor rather than the components themselves. The calculations
give

\begin{equation}
T_{t}^{t(q)}-T_{r}^{r(q)}=\frac{1}{96\pi ^{2}m^{2}}\beta z^{2}+{\cal O}%
\left( z^{3}\right) \text{,}
\end{equation}
where 
\begin{eqnarray}
&&\beta ^{(0)}=-\frac{1}{2520f_{0}^{2}\rho ^{6}r_{+}^{6}}%
(112A_{1}^{2}f_{0}^{2}\rho
^{4}+464A_{1}^{2}f_{0}^{2}r_{+}^{4}-704A_{2}f_{0}^{2}r_{+}^{5}  \nonumber \\
&&+200A_{1}f_{0}f_{1}r_{+}^{5}+209f_{1}^{2}r_{+}^{5}+64f_{0}f_{2}r_{+}^{6})+%
\frac{\eta }{30f_{0}^{2}\rho ^{6}r_{+}^{6}}(40A_{1}^{2}f_{0}^{2}\rho ^{4} 
\nonumber \\
&&-8A_{2}f_{0}^{2}\rho ^{4}r_{+}+2A_{1}f_{0}f_{1}\rho
^{4}r_{+}+80A_{1}^{2}f_{0}^{2}\rho
^{2}r_{+}^{2}+168A_{1}^{2}f_{0}^{2}r_{+}^{4}  \nonumber \\
&&+17f_{1}^{2}\rho ^{2}r_{+}^{4}-32f_{0}f_{2}\rho
^{2}r_{+}^{4}-552A_{2}f_{0}^{2}r_{+}^{5}+42A_{1}f_{0}f_{1}r_{+}^{5} 
\nonumber \\
&&+69f_{1}^{2}r_{+}^{6}-96f_{0}f_{2}r_{+}^{6})-\frac{\eta ^{2}}{%
2f_{0}^{2}\rho ^{6}r_{+}^{6}}(16A_{1}^{2}f_{0}^{2}\rho
^{4}+176A_{1}^{2}f_{0}^{2}\rho ^{2}r_{+}^{2}  \nonumber \\
&&-96A_{2}f_{0}^{2}\rho ^{2}r_{+}^{3}+24A_{1}f_{0}f_{1}\rho
^{2}r_{+}^{3}+144A_{1}^{2}f_{0}^{2}r_{+}^{4}-576A_{2}f_{0}^{2}r_{+}^{5}-24A_{1}f_{0}f_{1}r_{+}^{5}
\nonumber \\
&&+93f_{1}^{2}r_{+}^{6}-192f_{0}f_{2}r_{+}^{6})+\frac{6\eta ^{3}}{%
f_{0}^{2}\rho ^{6}r_{+}^{6}}(40A_{1}^{2}f_{0}^{2}\rho
^{4}-8A_{2}f_{0}^{2}\rho ^{4}r_{+}+2A_{1}f_{0}f_{1}\rho ^{4}r_{+}  \nonumber
\\
&&+64A_{1}^{2}f_{0}^{2}\rho ^{2}r_{+}^{2}-96A_{2}f_{0}^{2}\rho
^{2}r_{+}^{3}+24A_{1}f_{0}f_{1}\rho
^{2}r_{+}^{3}+40A_{1}^{2}f_{0}r_{+}^{4}+17f_{1}\rho ^{2}r_{+}^{4}  \nonumber
\\
&&-32f_{0}f_{2}\rho
^{2}r_{+}^{4}+104A_{2}f_{0}^{2}r_{+}^{5}+46A_{1}f_{0}f_{1}r_{+}^{5}-8f_{1}^{2}r_{+}^{6}+32f_{0}f_{2}r_{+}^{6})%
\text{,}
\end{eqnarray}
\begin{eqnarray}
\beta ^{(1/2)} &=&-\frac{1}{840f_{0}^{2}\rho ^{6}r_{+}^{5}}%
(56A_{2}f_{0}^{2}\rho ^{4}-14A_{1}f_{0}f_{1}\rho
^{4}+288A_{1}^{2}f_{0}^{2}r_{+}^{3}  \nonumber \\
&&-119f_{1}^{2}\rho ^{2}r_{+}^{3}+224f_{0}f_{2}\rho
^{2}r_{+}^{3}+96A_{2}f_{0}^{2}r_{+}^{4}-96A_{1}f_{0}f_{1}r_{+}^{4}  \nonumber
\\
&&+318f_{1}^{2}r_{+}^{5}-192f_{0}f_{2}r_{+}^{5})\text{,}
\end{eqnarray}
and 
\begin{eqnarray}
&&\beta ^{(1)}=-\frac{1}{420f_{0}^{2}\rho ^{6}r_{+}^{5}}(336A_{2}f_{0}^{2}%
\rho ^{4}-84A_{1}f_{0}f_{1}\rho ^{4}+280A_{1}^{2}f_{0}^{2}\rho
^{2}r_{+}+1680A_{2}f_{0}^{2}\rho ^{2}r_{+}^{2}  \nonumber \\
&&-420A_{1}f_{0}f_{1}\rho
^{2}r_{+}^{2}+1296A_{1}^{2}f_{0}^{2}r_{+}^{3}-714f_{1}^{2}\rho
^{2}r_{+}^{3}+1344f_{0}f_{2}\rho ^{2}r_{+}^{3}+1440A_{2}f_{0}^{2}r_{+}^{4} 
\nonumber \\
&&-1440A_{1}f_{0}f_{1}r_{+}^{4}+1935f_{1}^{2}r_{+}^{5}-2880f_{0}f_{2}r_{+}^{5})%
\text{,}
\end{eqnarray}
for the quantized massive scalar, neutral spinor, and vector field,
respectively.

\section{T$_{\protect\mu }^{\protect\nu (q)}$of massive fields in the
spacetime of extremal Reissner-Nordstr\"{o}m black hole}

Although the stress-energy tensor of the massive fields in the spacetime of
the extremal Reissner-Nordstr\"{o}m black hole may be easily constructed
taking extremality limits in the results of Refs. \cite{matrn} and \cite{ahs}%
, below we collect the formulas that have been used in the backreaction
calculations. For the quantized massive scalar field with an arbitrary
curvature coupling one has 
\begin{eqnarray}
T_{0}^{0(q)} &=&\frac{M^{2}\eta }{30240\pi ^{2}m^{2}r^{12}}%
\,(34398\,M^{4}-113904\,r\,M^{3}+139944\,M^{2}\,r^{2}-75600\,r^{3}\,M+15120%
\,r^{4})\,\,  \nonumber \\
&&+{\frac{M^{2}}{30240\pi ^{2}m^{2}r^{12}}}\,{(-1248\,M^{4}-45\,r^{4}+3084%
\,r\,M^{3}-2509\,M^{2}\,r^{2}+726\,r^{3}\,M)}\text{,}{\,}  \nonumber \\
&&
\end{eqnarray}
\begin{eqnarray}
T_{1}^{1(q)} &=&-\frac{M^{2}\eta }{30240\pi ^{2}m^{2}r^{12}}%
\,(4914\,M^{4}-21168\,r\,M^{3}+33432\,M^{2}\,r^{2}-23184\,r^{3}\,M+6048%
\,r^{4})\,\,  \nonumber \\
&&-{\frac{M^{2}}{30240\pi ^{2}m^{2}r^{12}}}\,{(-444\,M^{4}-477\,r^{4}+1932%
\,r\,M^{3}-2969\,M^{2}\,r^{2}+1950\,r^{3}\,M)\,}\text{,}  \nonumber \\
&&
\end{eqnarray}
and 
\begin{eqnarray}
T_{2}^{2(q)} &=&{\frac{M^{2}\eta }{30240\pi ^{2}m^{2}r^{12}}}\,{\
\,(44226\,M^{4}-143136\,r\,M^{3}+172536\,M^{2}\,r^{2}-91728\,r^{3}\,M+18144%
\,r^{4})\,}  \nonumber \\
&&-{\frac{M^{2}}{30240\pi ^{2}m^{2}r^{12}}}\,{\,(3066\,M^{4}-10356\,r%
\,M^{3}+12953\,M^{2}\,r^{2}-7086\,r^{3}\,M+1431\,r^{4})}\text{,}  \nonumber
\\
&&
\end{eqnarray}
wheras for the massive spinor field one obtains 
\begin{equation}
T_{0}^{0(q)}={\frac{M^{2}}{40320\pi ^{2}m^{2}r^{12}}}\,{(4917\,M^{4}-21496%
\,r\,M^{3}+32376\,r^{2}\,M^{2}-20080\,M\,r^{3}+4320\,r^{4})\,}\text{,}
\end{equation}
\begin{equation}
T_{1}^{1(q)}={\frac{M^{2}}{40320\pi ^{2}m^{2}r^{12}}}\,{(2253\,M^{4}-8680\,r%
\,M^{3}+12000\,r^{2}\,M^{2}-7120\,M\,r^{3}+1584\,r^{4})\,}\text{,}
\end{equation}
and 
\begin{equation}
T_{2}^{2(q)}=-{\frac{M^{2}}{40320\pi ^{2}m^{2}r^{12}}}\,{\,(9933%
\,M^{4}-23552\,M\,r^{3}+42888\,r^{2}\,M^{2}-33984\,r\,M^{3}+4752\,r^{4})}%
\text{.}
\end{equation}
Finally for the massive vector field in the extremality limit one has 
\begin{equation}
T_{0}^{0(q)}={\frac{M^{2}}{10080\pi ^{2}m^{2}r^{12}}}\,{(31057\,M^{4}-107516%
\,r\,M^{3}+135391\,M^{2}\,r^{2}-72690\,r^{3}\,M+13815\,r^{4})\,}\text{{,}}
\end{equation}
\begin{equation}
T_{1}^{1(q)}={\frac{M^{2}}{10080\pi ^{2}m^{2}r^{12}}}\,{(5365\,M^{4}-16996%
\,r\,M^{3}+19349\,M^{2}\,r^{2}-9398\,r^{3}\,M+1737\,r^{4})\,}\text{{,}}
\end{equation}
and 
\begin{equation}
T_{2}^{2(q)}=-{\frac{M^{2}}{10080\pi ^{2}m^{2}r^{12}}}\,{(13979\,M^{4}+5211%
\,r^{4}-26854\,r^{3}\,M+51789\,M^{2}\,r^{2}-44068\,r\,M^{3})}\text{.}
\end{equation}

\smallskip

\begin{table}[tbp]
\caption{The coefficients $c_{i}^{(s)}$ for the massive scalar, spinor, and
vector field. Note that to obtain the result for the massive neutral spinor
field one has to multiply $W_{ren}^{(1)}$ by the factor $1/2.$}
\label{table1}
\begin{tabular}{cccc}
& s = 0 & s = 1/2 & s = 1 \\ 
$c^{(s)}_{1} $ & ${\frac{1}{2}}\xi^{2}\,-\,{\frac{1}{5}} \xi $\,+\,${\frac{1%
}{56}}$ & $- {\frac{3}{140}}$ & $- {\frac{27}{280}}$ \\ 
$c^{(s)}_{2}$ & ${\frac{1}{140}}$ & ${\frac{1}{14}}$ & ${\frac{9 }{28}}$ \\ 
$c^{(s)}_{3}$ & $\left( {\frac{1}{6}} - \xi\right)^{3}$ & ${\frac{1}{432}}$
& $- {\frac{5}{72}}$ \\ 
$c^{(s)}_{4}$ & $- {\frac{1}{30}}\left( {\frac{1}{6}} - \xi\right) $ & $- {%
\frac{1}{90}}$ & ${\frac{31}{60}}$ \\ 
$c^{(s)}_{5}$ & ${\frac{1}{30}}\left( {\frac{1}{6}} - \xi\right)$ & $- {%
\frac{7}{720}}$ & $- {\frac{1}{10}}$ \\ 
$c^{(s)}_{6}$ & $- {\frac{8}{945}} $ & $- {\frac{25 }{378}}$ & $- {\frac{52}{%
63}}$ \\ 
$c^{(s)}_{7}$ & ${\frac{2 }{315}}$ & ${\frac{47}{630}}$ & $- {\frac{19}{105}}
$ \\ 
$c^{(s)}_{8}$ & ${\frac{1}{1260}}$ & ${\frac{19}{630}} $ & ${\frac{61}{140}} 
$ \\ 
$c^{(s)}_{9}$ & ${\frac{17}{7560}}$ & ${\frac{29}{3780}}$ & $- {\frac{67}{%
2520}}$ \\ 
$c^{(s)}_{10}$ & $- {\frac{1}{270}}$ & $- {\frac{1}{54}} $ & ${\frac{1}{18}}$%
\end{tabular}
\end{table}





%
%

%
%

\end{document}